\documentclass[aps,pra,floatfix,twocolumn,superscriptaddress,showpacs]{revtex4} 

\usepackage{graphicx}
\usepackage{graphics}
\usepackage{amssymb}
\usepackage{amsmath}
\usepackage{epsfig}
\usepackage{latexsym}
\usepackage{color}
\usepackage{subfigure}

\newcommand{\tr}{{\rm tr \thinspace}}
\newcommand{\bra}[1]{\left\langle{#1}\right\vert}
\newcommand{\ket}[1]{\left\vert{#1}\right\rangle}
\def\ketc[#1]{\vert #1 \rangle}
\def\brac[#1]{\langle #1 \vert}

\newcommand{\expect}[1]{\langle{#1}\rangle}

\newcommand{\beq}{\begin{equation}}
\newcommand{\eeq}{\end{equation}}
\newcommand{\bqa}{\begin{eqnarray}}
\newcommand{\eqa}{\end{eqnarray}}
\newcommand{\nn}{\nonumber}

\newcommand{\erf}[1]{Eq.~(\ref{#1})}
\newcommand{\dg}{^\dagger}

\begin{document}

\title{Adaptive homodyne phase discrimination and qubit measurement}

\author{Mohan Sarovar}
\email{msarovar@berkeley.edu} \affiliation{Department of Chemistry and Pitzer Center for Theoretical Chemistry, University of California, Berkeley, California 94720, USA}

\author{K. Birgitta Whaley}
\affiliation{Department of Chemistry and Pitzer Center for Theoretical Chemistry, University of California, Berkeley, California 94720, USA}


\begin{abstract}
Fast and accurate measurement is a highly desirable, if not vital, feature of quantum computing architectures. In this work we investigate the usefulness of adaptive measurements in improving the speed and accuracy of qubit measurement. We examine a particular class of quantum computing architectures, ones based on qubits coupled to well controlled harmonic oscillator modes (reminiscent of cavity-QED), where adaptive schemes for measurement are particularly appropriate. In such architectures, qubit measurement is equivalent to phase discrimination for a mode of the electromagnetic field, and we examine adaptive techniques for doing this. In the final section we present a concrete example of applying adaptive measurement to the particularly well-developed circuit-QED architecture. 
\end{abstract}
\pacs{03.67.-a, 03.67.Hk, 32.80.Qk}

\maketitle


\section{Introduction}
\label{sec:intro}
The exquisite control demanded by quantum information processing tasks over the initialization, evolution, and measurement of physical systems has motivated an intense level of study into these processes during the past decade. The numerous and diverse proposals for implementing quantum computing mean that we are learning how to control these processes in an array of physical systems of varying scale and complexity.

Several promising avenues for solid state realizations of quantum computers have been proposed (e.g. \cite{Kan-1998, Los.DiV-1998, Nak.Pas.etal-1999, Ima.Aws.etal-1999, Cor.Laf.etal-2000, Bla.Hua.etal-2004, Chi.Ber.etal-2004}). A recent trend in such solid state realizations, particularly ones based on superconducting circuits, has been to couple the qubits that form the quantum computer to well controlled harmonic modes of the electromagnetic (EM) field (e.g. circuit-QED \cite{Bla.Hua.etal-2004}, and Refs. \cite{Iri.Sch-2003, Gel.Cle-2005, Ata.Dre.etal-2007}). Such qubits-plus-harmonic-mode systems closely resemble the cavity-QED paradigm pioneered in quantum optics \cite{Mab.Doh-2002}, with atoms now being replaced by the qubits (which can be thought of \textit{artificial atoms}). Typically a system of this type is well described by the multi-atom Jaynes-Cummings (JC) model:
\beq
\mathcal{H}_{JC} = \sum_j \frac{\omega_a^j}{2}\hat{\sigma}_z^j + \omega_f\left(\hat{a}\dg\hat{a}+\frac{1}{2}\right) + \sum_j g_j(\hat{a}\hat{\sigma}_+^j + \hat{a}\dg\hat{\sigma}_-^j),
\eeq
where $\omega_a^j$ is the energy splitting of the $j$th qubit, $\omega_f$ is the frequency of the EM mode, $g_j$ is the coupling strength between qubit $j$ and the mode, $\hat{a}~ (\hat{a}\dg)$ is the annihilation (creation) operator for the field mode, and $\hat{\sigma}_+^j ~(\hat{\sigma}_-^j)$ is the raising (lowering) operator for the $j$th qubit. We have set $\hbar$ to 1 and will do so for the remainder of the paper. This Hamiltonian accounts for the coherent interaction between the elements; additional terms must be added to describe dissipation (spontaneous emission for the qubits and damping for the field mode), and driving of the field mode (e.g. see Ref. \cite{Bla.Hua.etal-2004}).

In the context of quantum computing, the cavity in such solid state cavity-QED systems has a dual function: as a bus that conveys quantum information between the qubits, and as an enabler of qubit measurement. The latter role comes from the fact that the well defined coupling between the qubit/s and the EM mode allows one to infer the state of the qubit/s from the state of the mode. In particular, when all the qubits are detuned from the cavity frequency, i.e. $\Delta_j \equiv | \omega_a^j - \omega_f |\gg g_j ~~~ \forall j$, then a measurement of the cavity is equivalent to a quantum non-demolition (QND) measurement of the state of the qubit/s. To see this, we note that in this detuned (off-resonant) limit, the dynamics of the system is well approximated by the dispersive Jaynes-Cummings Hamiltonian \cite{Ger.Kni-2005}:
\bqa
\label{eq:dispham}
\mathcal{H}_{disp} &=& \left(\omega_f + \sum_j \chi_j \hat{\sigma}_z^j\right) \hat{a}\dg\hat{a} + \frac{1}{2}\sum_j (\omega_a^j + \chi_j)\hat{\sigma}_z^j, \nn \\
\eqa
where $\chi_j \equiv g_j^2/\Delta_j$. The usual dispersive JC Hamiltonian has an additional term proportional to $~ \hat{\sigma}_+^i \hat{\sigma}_-^j + \hat{\sigma}_-^i \hat{\sigma}_+^j ~~~ \forall i,j ~~ i\neq j$ that describes the interqubit coupling enabled by virtual excitations of the cavity. This term is reduced in magnitude when the qubits are made off-resonant with \textit{each other} \cite{Bla.Gam.etal-2007}, and since this is the regime that is most useful for measurement, we will ignore this mutual interaction term in the dispersive Hamiltonian. Note that the first term in \erf{eq:dispham} describes a ``pull" of cavity mode frequency by the qubit/s, and that this pull is dependent on the state of the qubit/s (in the $\hat{\sigma}_z$ basis). Intuitively then, if all the $\chi_j$ are different (i.e. all the $\omega_a^j$ are different), there is a unique cavity pull for each state of the qubit/s (in fact, it is not sufficient that all the pulls simply be different for a unique determination; see section \ref{sec:largen} for the complete condition). This pull is exactly what is used to measure the state of the qubit/s using the cavity by virtue of the fact that a measurement of the phase of the cavity output field determines this frequency pull \cite{Bla.Hua.etal-2004, Sar.Goa.etal-2005}. For example, when we have one qubit, the dispersive JC Hamiltonian takes the form $H = \left(\omega_f + \chi \hat{\sigma}_z\right) \hat{a}\dg\hat{a} + \frac{1}{2}(\omega_a + \chi)\hat{\sigma}_z$, and a probe coherent state of the cavity undergoes the following input-ouput map:
\bqa
\ket{\alpha} \rightarrow \left\{ \begin{array}{ll} \ket{\alpha e^{i\phi}} & \textrm{if } \expect{\hat{\sigma}_z}=1 \\ \ket{\alpha e^{-i\phi}} & \textrm{if } \expect{\hat{\sigma}_z}=-1 \end{array} \right. ,
\eqa
where $\phi = \tan^{-1}\left( \frac{g^2}{\kappa \Delta}\right)$ is the magnitude of the phase change (here, $\kappa$ is the decay rate of the cavity).

\begin{figure}
\includegraphics[scale=0.45]{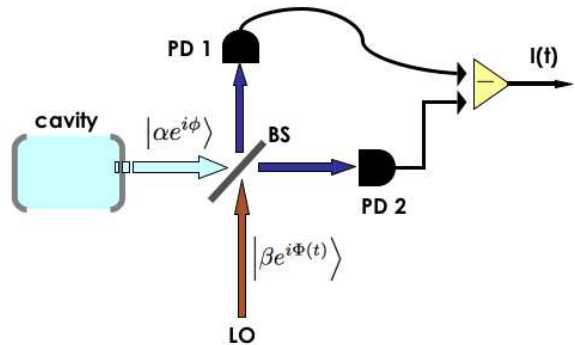}
\caption{(Color online) The typical method for measuring the phase of a field mode: a \textit{dyne} measurement. The input mode (for the measurement) is the output of the cavity, and in the cavity QED scenario we are considering, it will have an unknown phase ($\phi$) due to the pull of qubit/s in the cavity. The local oscillator (LO) phase, $\Phi(t)$, is constant for a homodyne measurement and rapidly oscillating for a heterodyne measurement.} \label{fig:dyne}
\end{figure}

Therefore, in such qubit/s-plus-oscillator systems a measurement of the state of the qubit/s is often a phase measurement of the oscillator, and the task of qubit/s state determination becomes the same as that of phase discrimination for a harmonic oscillator mode. Thus, if we are interested in improving qubit measurement, we are naturally led to the question: how best to realize phase discrimination of an EM mode? Quantum state discrimination, especially in the optical context, is a well studied problem with a long history mostly motivated by communication problems \cite{Hel-1976}. The standard method for phase measurement of a field mode is to perform a \textit{dyne} measurement of the mode \cite{wallsandmilburn, Gar.Zol-2004}. This involves mixing the mode with a local oscillator on a beam-splitter and performing photodetection on the two output ports of the beamsplitter; see Figure \ref{fig:dyne}. A \textit{homodyne} measurement uses a fixed local oscillator phase and thus measures one particular quadrature of the input mode, while a \textit{heterodyne} measurement uses a rapidly varying local oscillator phase, which results in a measurement that samples all quadratures of the input mode equally. Both these dyne measurements are static schemes -- they do not use information gained about the unknown phase to update the local oscillator phase in order to optimize further information gain. Wiseman \cite{Wis-1995} has shown that adaptive schemes can outperform such static schemes for the task of phase \textit{estimation} for field modes. This was followed by a study by Bargatin \cite{Bar-2005} showing that general coherent state discrimination is improved by adaptive schemes. While Refs. \cite{Wis-1995} and \cite{Bar-2005} use different measures of performance, both clearly show an improvement with adaptive strategies. Also, Dolinar showed already in 1973 \cite{Dol-1973} that the discrimination of coherent states by photon counting can be done more efficiently by using adaptive measurements and this was recently confirmed with both simulation and experiment by Geremia \textit{et. al.} \cite{Ger-2004, Coo.Mar.etal-2007}. In fact, the adaptive strategy detailed in \cite{Dol-1973} achieves the optimally discriminating measurement in the limit of infinitely fast feedback. These observations motivate the central question of this paper: can one \textit{distinguish} optical phases, and ultimately qubit states in the cavity-QED context, more efficiently using adaptive dyne measurements? The results of Refs. \cite{Wis-1995, Bar-2005, Dol-1973} would suggest so, and we explore the question in detail in the following sections.

Firstly, in section \ref{sec:back} we provide some background to adaptive dyne measurements. Then, in section \ref{sec:phase_disc} we examine the task of adaptive phase discrimination, first looking at the case of two phases (i.e. measuring a single qubit in the cavity QED quantum computing scenario) and then at adaptive phase discrimination for an arbitrary number of phases. In that section, we treat in detail an example of adaptive state discrimination in circuit-QED \cite{Bla.Hua.etal-2004}, a canonical solid-state implementation of the qubits-plus-harmonic-mode system. We conclude in section \ref{sec:conc} with a summary of the results.

\section{Adaptive dyne measurements}
\label{sec:back}
A typical (balanced) dyne measurement is illustrated in Fig. \ref{fig:dyne}. The field mode with the unknown phase ($\phi$) is mixed with a mode in a coherent state of known phase ($\Phi(t)$) and large amplitude ($\beta$), and both outputs are measured by photodetectors. The output signal, $I(t)$, also known as the photocurrent, in the interval $[t, t+\delta t)$ is defined as \cite{Wis.Mil-1993b, Wis.Mil-1993}:
\bqa
\label{eq:sig}
I(t) &=& \lim_{\delta t \rightarrow 0} \lim_{|\beta | \rightarrow \infty} \frac{\delta N_2(t) - \delta N_1(t)}{|\beta | \delta t} \nn \\
&=& e^{-t/2}\expect{\hat{a}e^{-i\Phi(t)} + \hat{a}\dg e^{i\Phi(t)}} + \xi(t),
\eqa
where $\delta N_i(t)$ is the photocount at the $i$th detector, $\hat{a}, \hat{a}\dg$ are the annihilation and creation operators for the input mode (with unknown phase), and $\xi(t)$ is delta-correlated Gaussian white noise that captures the unavoidable shot noise of the detectors \cite{Gar.Zol-2004}. The decaying exponential envelope in the above average signal reflects the fact that the probe beam is a finite pulse (we do not consider continuous probe beams). Note that by varying $\Phi(t)$ we can measure any quadrature of the input mode -- for example, $\Phi(t) = 0$ measures the $\hat{x} \propto \hat{a}+\hat{a}\dg$ quadrature. As mentioned above, a rapidly varying $\Phi(t)$ samples all quadratures of the input mode equally. In the case that the input mode is a coherent state, $\ket{\alpha e^{i\phi}}$, where $\alpha$ is real, then the output photocurrent simplifies to
\beq
\label{eq:sig2}
I(t) = 2\alpha e^{-t/2} \cos[\Phi(t) - \phi] + \xi(t).
\eeq
An equivalent quantity that will prove to be more useful in the following is the photocurrent increment
\beq
\label{eq:sig3}
I(t)dt = 2\alpha e^{-t/2} \cos[\Phi(t) - \phi]dt + dW(t),
\eeq
where $dW(t)$ are Wiener increments of mean zero and variance $dt$ \cite{Gar.Zol-2004}. Note that the probe signal magnitude $\alpha$ functions as an effective signal-to-noise (SNR) parameter.

\begin{figure}
\includegraphics[scale=0.42]{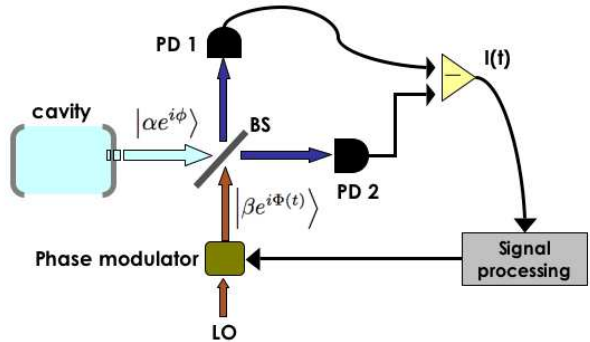}
\caption{(Color online) An adaptive scheme for measuring the phase of a field mode. The local oscillator phase is dynamically modified according to the measured values of the photocurrent. \label{fig:adap_dyne}}
\end{figure}

Now, consider the modification to the standard dyne measurement shown in Fig. \ref{fig:adap_dyne}. Here, the information contained in the output photocurrent is used to dynamically modify the phase of the local oscillator. This introduces a feedback loop which opens up the possibility of adapting the dyne measurement according based on information received in earlier stages of the measurement. 

Wiseman first considered such an adaptive scheme for the task of phase estimation. He showed that using an adaptive homodyne scheme one can reduce the variance of the estimate faster that with any static scheme \cite{Wis-1995}. His result can be understood simply by noting that the sensitivity of a dyne measurement, defined as $| \delta I(t) / \delta \phi |$, is greatest when $| \sin[\phi - \Phi(t)]|$ is maximized. This can be achieved by a $\Phi(t)$ such that $\phi-\Phi(t) = \pm \pi/2$; that is, by a local oscillator that is in quadrature with the unknown phase. Wiseman's scheme uses this fact to choose the local oscillator phase $\Phi(t) = \hat{\phi}(t)+\pi/2$ where $\hat{\phi}(t)$ is the current estimate of the phase $\phi$. This estimate is formed from the measurement current up till time $t$. Thus as the phase estimate improves, the sensitivity of the measurement increases.

We want to use this idea of adaptive measurement to \textit{discriminate} between a given number of optical phases. 

\section{Adaptive phase discrimination}
\label{sec:phase_disc}
\begin{figure}
\includegraphics[scale=0.5]{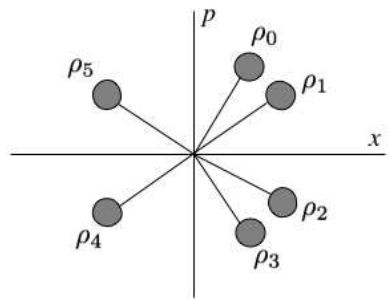}
\caption{A phase space diagram showing the possible phases in a phase discrimination task with 6 hypotheses. \label{fig:ho_n}}
\end{figure}

The analysis of phase discrimination can be made quantitative by treating it as a formal \textit{decision} process, and more particularly, as a hypothesis testing task. Each possible phase forms a particular hypothesis and the task is to determine the correct phase from the experimental observations: the photocurrent up till time $t$, which we denote by $\mathbf{I}_{[0,t)}$. More precisely, the hypotheses are $\mathcal{H}_j : \rho = \rho_j \equiv \ket{\alpha_j}\bra{\alpha_j} = \ket{\alpha e^{i\varphi_j}}\bra{\alpha e^{i\varphi_j}}$, one for each possible phase, $\varphi_j$, resulting from a different pull of the probe beam frequency by the qubits in the cavity. We restrict ourselves to probe beams that are coherent states and linear cavity dynamics (so the output coherent state amplitude is the same for all qubit states -- only the phase changes) since this is experimentally most relevant. We will comment on non-linear cavity dynamics in the final section of the paper. Figure \ref{fig:ho_n} shows the coherent state ``constellation" associated with a phase discrimination task; the members of the constellation form the hypotheses.

By the \textit{likelihood principle} \cite{Ber-1980}, all the information relevant to the decision is contained in the likelihood functions for the hypotheses, defined as:
\beq
\label{eq:lik}
\mathcal{L}_j \equiv \textrm{Pr}(\mathbf{I}_{[0,t)} | \mathcal{H}_j, \Phi_{[0,t)} ) = \textrm{Pr}( \mathbf{I}_{[0,t)} | \rho=\rho_j, \Phi_{[0,t)} ).
\eeq
As the right-hand side indicates, the likelihood function is a conditional probability for the observed measurement record given the particular hypothesis. We also condition on the local oscillator phase history in anticipation that this will become a dynamical variable. The likelihood principle asserts that all decisions regarding the hypotheses are to be made from these likelihood functions and therefore we are led to consider them in more detail \footnote{For a detailed discussion about using the likelihood principle in decision problems see Chapter 1 of Ref. \cite{Ber-1980}.}. 

Using Bayes' rule, we express the \textit{a posteriori} probabilities of various hypotheses in terms of the likelihood functions as:
\bqa
\label{eq:cond_prob}
\mathcal{P}_j \equiv  \textrm{Pr}( \rho=\rho_j | \mathbf{I}_{[0,t)}, \Phi_{[0,t)} ) &=& \frac{\textrm{Pr}( \mathbf{I}_{[0,t)} | \rho=\rho_j, \Phi_{[0,t)}  ) \zeta_j }{ \textrm{Pr}( \mathbf{I}_{[0,t)} | \Phi_{[0,t)}) } \nn \\
&=& \frac{\mathcal{L}_j \zeta_j}{\mathcal{N}},
\eqa
where $\zeta_j$ is the prior probability that $\rho=\rho_j$, and the denominator is a normalizing quantity: $\mathcal{N} \equiv \textrm{Pr}(\mathbf{I}_{[0,t)} | \Phi_{[0,t)} ) = \sum_i \textrm{Pr}(\mathbf{I}_{[0,t)} | \rho=\rho_i, \Phi_{[0,t)} ) \zeta_i = \sum_i \mathcal{L}_i \zeta_i$. The prior probabilities are assumed to be known (they are the prior probabilities of the qubit/s states), and from here onwards we will assume that all qubit states are equally likely and therefore use a uniform prior. In this case, note that the \textit{a posteriori} probabilities for the hypotheses ($\mathcal{P}_j$) are simply normalized versions of the likelihood functions ($\mathcal{L}_j$). Now, quantum measurement theory tells us that 
\bqa
\textrm{Pr}( \mathbf{I}_{[0,t)} | \rho=\rho_j, \Phi_{[0,t)}  ) = \tr(\rho_j \hat{F}_t), 
\eqa
where $\hat{F}_t$ is the POVM (parameterized by the continuous variable $t$) that corresponds to the observed photocurrent, $\mathbf{I}_{[0,t)}$, and local oscillator history, $\Phi_{[0,t)}$. Wiseman has evaluated this POVM as being \cite{Wis-1996,Wis-1995}:
\bqa
\label{eq:povm}
\hat{F}_t \equiv \hat{F}(R_t, S_t) &=& P_0(R_t, S_t) \hat{G}_t(R_t, S_t)  \nn \\
&=& P_0(R_t, S_t) \exp\left(\frac{1}{2} S_t \hat{a}^{\dagger 2} + R_t \hat{a}\dg \right)\nn
\\ &&  \times \exp(-\hat{a}\dg \hat{a} t)\exp\left(\frac{1}{2} S_t^* \hat{a}^2  + R_t^* \hat{a} \right), \nn \\
\eqa
where $R_t$ and $S_t$ are functionals of $\mathbf{I}_{[0,t)}$ that are \textit{sufficient statistics} \cite{Ber-1980} which completely capture the influence of the measurement record. They are explicitly:
\bqa
R_t &\equiv& R_t[\mathbf{I}_{[0,t)}] = \int_0^t e^{i \Phi(s)} e^{-s/2} I(s) ds \nn \\
S_t &\equiv& S_t[\mathbf{I}_{[0,t)}] = -\int_0^t e^{i2\Phi(s)} e^{-s} ds.
\eqa
Note that the second quantity is also a functional of the measurement record if the local oscillator phase, $\Phi(t)$, depends on the measurement record up till time $t$, $\mathbf{I}_{[0,t)}$, as will be the case in our adaptive scheme. $P_0(R_t, S_t)$ in \erf{eq:povm} is a normalizing factor which is inconsequential to us because it is present in both the numerator and denominator of the expression for the \textit{a posteriori} probabilities for the hypotheses, \erf{eq:cond_prob}, which are ultimately what we are interested in. Thus we redefine the likelihood functions as:
\beq
\label{eq:lj}
\mathcal{L}_j = \bra{\alpha e^{i\varphi_j}}\hat{G}_t \ket{\alpha e^{i\varphi_j}},
\eeq
and evaluate the expectation value to obtain:
\begin{widetext}
\bqa
\label{eq:lj_simp}
\mathcal{L}_j(t) &=& \exp\{- |\alpha |^2(1-e^{-t})\}  \exp \left\{ \mathbf{Re}(S_t \alpha_j^{* 2}) + 2\mathbf{Re}(R_t \alpha_j^{*})\right\} \nn \\
&=& \mathcal{C}_t \exp \left\{ -2|\alpha |\int_0^t (|\alpha | e^{-s}\cos^2(\Phi(s) - \varphi_j) - e^{-s/2}\cos(\Phi(s) - \varphi_j) I(s) )~ds  \right\},
\eqa
where $\mathcal{C}_t \equiv \exp\{- |\alpha |^2(1-e^{-t})\}$ is a prefactor that is common to all $j$. 
\end{widetext}

\erf{eq:lj_simp} gives us explicit expressions for the likelihood functions. We will utilize these to form the decision rules for the hypothesis testing problem in the following subsections. Note that the likelihood functions are themselves random because they are functions of the random observation process. 

\subsection{Adaptive phase discrimination between $N=2$ phases}
\label{sec:n2}
Consider the simplest case of discriminating between two possible phases. In the cavity-plus-qubits experimental setup this corresponds to having one qubit coupled to the cavity mode. In such a case we have two hypotheses: $\mathcal{H}_+: \rho = \rho_+ \equiv \ket{\alpha e^{i\varphi}}\bra{\alpha e^{i\varphi}}$ and $\mathcal{H}_-: \rho = \rho_- \equiv \ket{\alpha e^{-i\varphi}}\bra{\alpha e^{-i\varphi}}$, for some $\varphi$. In a two-hypothesis test, the decision policy is simple and obvious (at least once the likelihood principle has been adopted). It is based on the likelihood ratio, $\Lambda = \mathcal{L}_+ / \mathcal{L}_-$, which is simply the ratio between the likelihood functions for the two hypotheses. The decision policy is to accept $\mathcal{H}_+$ if $\Lambda > 1$ and $\mathcal{H}_-$ if $\Lambda < 1$ (in the degenerate case of $\Lambda=1$, a random decision can be made) \footnote{A strict statistician would state this decision policy in terms of \textit{rejecting} certain hypotheses -- since data can only be used in a statistical sense to reject hypotheses -- but for simplicity, we will ignore this distinction here.}.

Using the expressions for the likelihood functions derived above, and assuming equal priors (i.e. both states of the qubit are equally likely) we write the likelihood ratio as:
\begin{widetext}
\bqa
\label{eq:lambda_2}
\Lambda(t) &=& \exp \left\{ -2|\alpha |^2 \int_0^t e^{-s} [\cos^2(\Phi(s)-\phi) - \cos^2(\Phi(s) + \phi)]~ds \right. \nn \\
&& ~~~~~~~~~~~ \left. + 2|\alpha | \int_0^t e^{-s/2} [\cos(\Phi(s)-\phi) - \cos(\Phi(s)+\phi)] I(s) ~ds  \right\}.
\eqa
\end{widetext}
Because of the exponential in this expression, we will work with the log of the likelihood ratio; this merely changes the decision policy by changing the threshold (i.e., $\ln \Lambda(t) > 0$ or $\ln \Lambda(t) < 0$).

Given this decision policy, we can attempt to optimize the probability of a correct decision by using an adaptive scheme. That is, we want to choose $\Phi(t)$ at time $t$ such that the probability of a correct decision is maximized. This $\Phi(t)$ can depend on the photocurrent up till time $t$: $\mathbf{I}_{[0,t)}$. The probability we want to maximize is:
\bqa
p_c(t) &=& \frac{1}{2} \big[ \textrm{Pr}(\ln \Lambda(t)>0 ~|~ \rho = \rho_+) \nn \\
&& ~~~~~ + \textrm{Pr}(\ln \Lambda(t)<0 ~|~ \rho = \rho_-) \big] ,
\eqa
where again, we have assumed equal prior probabilities for the two phases. This optimization maximizes the probability of a correct decision at all times $t$. An alternative would be to maximize this probability only at some fixed time $T$, however, we shall find the former easier to do and of course, it implies the latter optimum. 

Using the expression for the measurement current, \erf{eq:sig3}, we can write this success probability as:
\begin{widetext}
\bqa
\label{eq:pc_expl}
2 p_c(t) &=& \textrm{Pr}\bigg( \int_0^t e^{-s/2}[\cos(\Phi(s)-\phi) - \cos(\Phi(s)+\phi)]~dW(s) > -|\alpha| \int_0^t e^{-s}~ds [\cos(\Phi(s)-\phi) - \cos(\Phi(s)+\phi)]^2 \bigg) \nn \\
&& + \textrm{Pr}\bigg( \int_0^t e^{-s/2}[\cos(\Phi(s)-\phi) - \cos(\Phi(s)+\phi)]~dW(s) < |\alpha| \int_0^t e^{-s}~ds [\cos(\Phi(s)-\phi) - \cos(\Phi(s)+\phi)]^2 \bigg). \nn \\
\eqa
\end{widetext}
Note that the left hand sides of both probability arguments are the same; namely, some random variable that is a sum of weighted $dW(s)$ terms. The sum is not necessarily Gaussian because the $\Phi(s)$ could depend on $dW(s'), s'<s$. However, the left-hand-side can simply be treated as a random variable, which we will label $\mathbb{X}(t)$. Hence the success probability takes the form $p_c(t) = 1/2[ \textrm{Pr}(\mathbb{X}(t)>-\chi(t)) + \textrm{Pr}(\mathbb{X}(t)<\chi(t)) ]$, where $\chi \equiv |\alpha|\int_0^t e^{-s}~ds[\cos(\Phi(s)-\phi) - \cos(\Phi(s)+\phi) ]^2$. Regardless of the distribution of $\mathbb{X}(t)$ (as long as the domain of the random variable is the entire real number line, which it is in this case) this probability will be maximized when $\chi(t)$ is maximized. Furthermore, maximizing $\chi(t)$ is trivial since it is an integral of a positive semidefinite quantity: $\chi(t)$ is maximized when the integrand is maximized for all $s$. And it is easy to show that $e^{-s}[\cos(\Phi(s)-\phi) - \cos(\Phi(s)+\phi) ]^2$ is maximized by the choice $\Phi(s) = \pi/2$ for all $s$.

\begin{figure}
\includegraphics[scale=0.5]{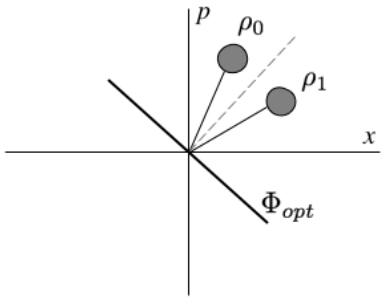}
\caption{A phase space diagram showing the possible phases in a phase discrimination task with two hypotheses. As shown in the main body, the optimal local oscillator phase is static and in quadrature with the bisector of the two candidate phases. \label{fig:ho_2}}
\end{figure}

Hence we see that the optimal strategy for distinguishing two phases is not an adaptive strategy at all, but rather a static one that keeps the local oscillator phase fixed at $\pi/2$. The above calculation can be generalized easily to the case of distinguishing two arbitrary phases $\varphi_0$ and $\varphi_1$ that do not necessarily add to zero: the optimal strategy then is $\Phi(t) = \pi/2 + (\varphi_0+\varphi_1)/2$. The uselessness of an adaptive strategy in this case can be understood by looking at the optimal dyne measurement on the phase space diagram of Fig. \ref{fig:ho_2}. The optimal measurement is such that the local oscillator phase is in quadrature with the bisector of the phases that are to be distinguished -- we will refer to this measurement as one that is \textit{symmetrically in quadrature}. Such a measurement gathers the most information relevant to the discrimination task per unit time. And since the phase space is two dimensional, this measurement exists for any two phases $\varphi_0$ and $\varphi_1$. This agrees with the recent result in Ref. \cite{Tak.Sas-2007} that the optimal Gaussian strategy (of which dyne measurements are a subset) for discriminating two coherent state phases is a static homodyne measurement. 

However, if the task is to distinguish between \textit{more} than two phases, such a symmetrically in quadrature measurement no longer exists. Therefore in such cases, we might expect an adaptive strategy to outperform a static one. We will investigate this in the following subsection.

\subsection{Adaptive phase discrimination between $N>2$ phases}
\label{sec:ng2}
When the number of phases to discriminate is greater than two, the problem is one of multiple hypothesis testing. This is a much harder problem and we do not have results of optimality here. A sensible decision policy for multiple hypothesis testing, when the likelihood principle has been adopted and the losses are uniform, is the maximum likelihood decision -- i.e., choose the hypothesis with the maximum likelihood function \cite{Ber-1980}. Given this policy, one might formulate a performance function for the decision problem as:
\beq
\mathcal{L}_{corr}(t) - \sum_{i\neq corr} \mathcal{L}_i(t)
\eeq
where $\mathcal{L}_{corr}$ is the likelihood of the correct phase. Then the task would be maximize this performance function by a choice of $\Phi_{[0,t)}$. However, note that the likelihood functions are non-differentiable functions and therefore many of the standard tools of optimization, such the calculus of variations or dynamic programming, become unsuitable for use.

Given this situation, we will use the results of the last subsection to motivate the following ad-hoc adaptive strategy for modulating the local oscillator phase:
\beq
\label{eq:adapt}
\Phi(t) = \frac{\pi}{2} + \frac{\varphi_M(t) + \varphi_m(t)}{2},
\eeq
where $\varphi_M(t)$ and $\varphi_m(t)$ are the phases with the largest and second-largest likelihood function at time $t$. This strategy involves keeping track of the likelihood functions for all the candidate phases (all hypotheses) and choosing to perform the symmetrically in quadrature measurement of the two hypotheses of largest likelihood. There are no claims of optimality for this strategy, but it is a sensible one to adopt given the decision rule. In the following, we will numerically evaluate this adaptive scheme in two illustrative situations. 

\subsubsection{Example 1: a circuit-QED example}
\label{sec:cqed}
The first numerical study is motivated by the circuit-QED architecture for quantum computation \cite{Wal.Sch.etal-2004,Bla.Hua.etal-2004}. This architecture is a direct analogy of the cavity-QED setups in quantum optics and therefore the discussion in the Introduction applies exactly. We consider the situation where there are two qubits coupled to the strip-line microwave resonator that functions as the cavity \cite{Wal.Sch.etal-2004}. We assume these qubits have different energy splittings and therefore will pull the cavity frequency by different amount when the cavity is detuned from both of them. Therefore, a measurement of the phase of the cavity output mode could yield four different values, each of which forms a hypothesis:
\beq
\phi = \left\{ \begin{array}{ll}
\varphi_0 \equiv \phi_1 + \phi_2 & \textrm{if } \expect{\sigma_z^1}=1, \expect{\sigma_z^2}=1 \\
\varphi_1 \equiv \phi_1 - \phi_2 & \textrm{if } \expect{\sigma_z^1}=1, \expect{\sigma_z^2}=-1 \\
\varphi_2 \equiv - \phi_1 + \phi_2 & \textrm{if } \expect{\sigma_z^1}=-1, \expect{\sigma_z^2}=1 \\
\varphi_3 \equiv -\phi_1 - \phi_2 & \textrm{if } \expect{\sigma_z^1}=-1, \expect{\sigma_z^2}=-1
\end{array} \right.
\eeq
where $\phi_i = \tan^{-1}(g_i^2/\kappa \Delta_i)$, and $g_i$ and $\Delta_i$ are, respectively, qubit $i$'s coupling strength to, and detuning from, the cavity mode ($\kappa$ is the cavity damping rate, which is a function of the capacitive coupling of the stripline to the output port). A wide range of values for $g_i$ are experimentally accessible in the circuit-QED experiments. Using the cavity pull reported in Ref. \cite{Bla.Hua.etal-2004} ($g^2/\kappa\Delta \approx 2.5$) as a guide, we set $\phi_1 = 4\pi/10, \phi_2 = 3\pi/10$.

\begin{figure}[t!]
      \subfigure[ ~Time evolution of the $\overline{\mathcal{P}_{corr}}$ for an example fixed probe signal magnitude (i.e. fixed SNR, $\alpha$).]{ \includegraphics[scale=0.25]{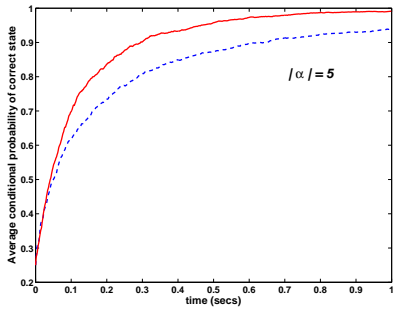} } \quad
       \subfigure[ ~The $\overline{\mathcal{P}_{corr}}$ as a function of probe signal magnitude (SNR) $\alpha$ at two fixed times.   The lines are interpolations between data points at integer $|\alpha |$ values to aid the eye. See main text for comments on the error bars.]{\includegraphics[scale=0.25]{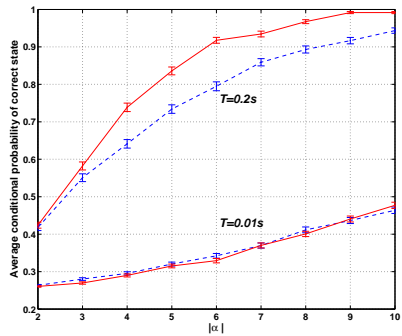} } \quad
      \subfigure[ ~The time taken for the $\overline{\mathcal{P}_{corr}}$ to reach 0.5 as a function of probe signal magnitude (SNR) $\alpha$. The lines are interpolations between data points at integer $|\alpha |$ to aid the eye.]{\includegraphics[scale=0.25]{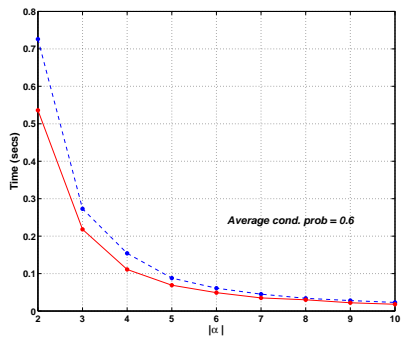} }
    \caption{(Color online) Figures contrasting the evolution of the average \textit{a posteriori} probability of the correct hypothesis ($\overline{\mathcal{P}_{corr}}$) under the adaptive and static schemes described in section \ref{sec:cqed}. In all the figures, the solid (red) line shows the evolution under the adaptive scheme and the dashed (blue) line shows evolution under the static heterodyne scheme.}
    \label{fig:cqedres}
\end{figure}

We now compare the adaptive strategy described by \erf{eq:adapt} to the static strategy of heterodyne detection. Static heterodyne detection, where $\Phi$ is cycled rapidly, is known to be superior to the alternative static strategy of homodyne detection when the number of phases to discern is $> 2$. 
The average performance of a static heterodyne scheme is linked to how fast the phase is cycled: it improves as the phase is cycled faster but the improvement plateaus after a threshold is passed. We determined this threshold and operated the static heterodyne scheme above it (numerically, at $100\pi$ rads/sec). We then simulate the measurement current generated by a given phase and track the likelihood function of the correct phase under the two different strategies. This is done 500 times in order to capture the average behavior and the results are shown in Fig. \ref{fig:cqedres}. 

\begin{figure}[h]
\includegraphics[scale=0.3]{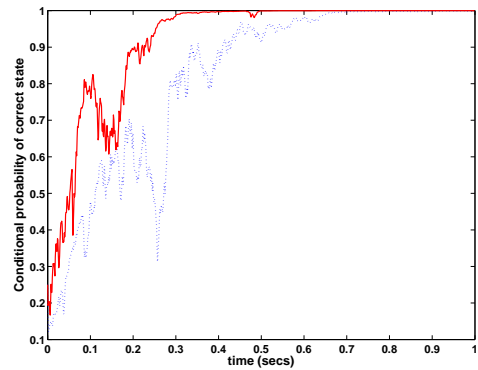}
\caption{(Color online) Example of a single run of the measurement in the circuit QED example of section \ref{sec:cqed}. The red (solid) line is the evolution of the \textit{a posteriori} probability of the correct state ($\mathcal{P}_{corr}$) under the adaptive scheme and the blue (dotted) line is its evolution under the static heterodyne scheme. \label{fig:res_single}}
\end{figure}

These figures show the average development of the \textit{a posteriori} probability for the correct hypothesis, $\overline{\mathcal{P}_{corr}}$, as a function of time and signal-to-noise, SNR (recall that this is measured by $\alpha$). $\mathcal{P}_{corr}$ is of course one of the $\mathcal{P}_j$ defined in \erf{eq:cond_prob}, and for the simulation we generated a measurement signal consistent with the smallest phase, $\varphi_1 \equiv \phi_1 - \phi_2 = \pi/10$. Hence, $\mathcal{P}_{corr} = \mathcal{P}_1$ in this case. The results indicate that, on average, the adaptive strategy outperforms the static strategy for intermediate times. For very short times, the static heterodyne scheme performs slightly better or equally well, because the adaptive scheme is ineffective at short times while the Bayesian filter accumulates initial data. And at long times, the performance of the two schemes becomes similar, although the adaptive strategy converges to large likelihood values faster. However, at intermediate times, figure \ref{fig:cqedres} shows that the adaptive strategy performs better in that: (i) it requires less time for the correct \textit{a posteriori} probability to reach any threshold value, and (ii) it requires a smaller signal-to-noise ratio to achieve the same level of certainty. Similar conclusions result from comparing the adaptive strategy with a static homodyne detection scheme, with the differences being slightly more pronounced. It should be noted that these graphs show average results; in contrast, a single run of the measurement would result in an evolution of the observer's state of knowledge as illustrated in Fig. \ref{fig:res_single}. Note that in Fig. \ref{fig:cqedres}(b) the error bars at each data point show the standard error of the mean values plotted, i.e., $\sigma /\sqrt{n}$, where $\sigma$ is the standard deviation and $n$ is the sample size (these error bars are omitted in Fig. \ref{fig:cqedres}(a) in the interest of clarity). The small errors bars are assurance that the sample size of 500 is sufficient to capture the average behavior well.

Figure \ref{fig:cqedres} shows the results for a particular choice of correct state. The average behavior of the adaptive scheme when averaged over each of the four possible phase states taken as the correct one, and the corresponding average probabilities from static heterodyne detection are illustrated with several examples in Table \ref{tab:av_cqed}. The table shows the average conditional probability of the correct state for $|\alpha |=5$ and for two choices of time (note that there are two averages taken here, one over many runs of the task -- which results in $\overline{\mathcal{P}_{corr}}$ -- and then another over the four possibilities for correct phase state).  The table clearly shows the advantage of the adaptive scheme, with increased probabilities of success at short and long times.

\begin{table}[htdp]
\begin{center}
\begin{tabular}{|c|c|c|c|}
\hline
Time (secs) & Detection scheme & Average $\overline{\mathcal{P}_{corr}}$ & $\sigma$ \\
\hline \hline
0.2 & Static & 0.8595 & 0.0164  \\
0.2 & Adaptive & 0.9089 & 0.0144 \\
1 & Static & 0.9680 & 0.0078 \\
1 & Adaptive & 0.9951 & 0.0026 \\
\hline
\end{tabular} 
\caption{$N=4$ detection: values of the average \textit{a posteriori} probability of the correct phase, $\overline{\mathcal{P}_{corr}}$, and standard deviation of this quantity, $\sigma$, at two time values, one short and one long, for fixed SNR ($|\alpha | = 5$).  \label{tab:av_cqed}}
\end{center}
\end{table}

\subsubsection{Example 2: large $N$ hypothesis testing}
\label{sec:largen}
As the number of phases to be discriminated, $N$, gets larger, the discrimination task approaches the estimation task considered by Wiseman in Ref. \cite{Wis-1995}. However as the number of qubits in the cavity increases, the conditions on the phase pull that each qubit induces ($\phi_i =  \tan^{-1}\left( \frac{g_i^2}{\kappa \Delta_i}\right)$) become stricter if we demand a unique determination of the state of the qubits from the phase measurement. It is not sufficient that each phase pull be different, but rather an additional condition is required for unique determination. This is, that no phase can equal the sum of any of the others (mod $2\pi$). For large $N$, this could conceivably be a difficult constraint, and could make such a cavity-mediated qubit measurement scheme demanding. 

\begin{figure}[h]
\includegraphics[scale=0.3]{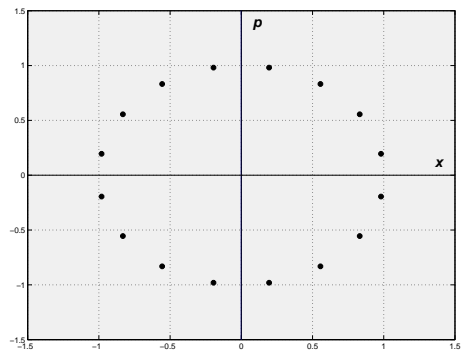}
\caption{The phase constellation (for $|\alpha | = 1$) for the discrimination task described in section \ref{sec:largen}. (The dots simply show the centers of the coherent states, the uncertainties in phase space are not shown). \label{fig:16const}}
\end{figure}

In this subsection, we assume that this constraint can be met and numerically examine the performance of the adaptive scheme described above for an example constellation with a large number of phases to distinguish. We consider four qubits coupled to the cavity mode, and thus have $16$ different phases to distinguish. The individual phase pulls of each qubit are: $\phi_1 = \pi/16, \phi_2 = \pi/8, \phi_3 = \pi/4, \phi_4 = \pi/2$. The $16$-element constellation is shown in Fig. \ref{fig:16const}. 

\begin{figure}[h]
\includegraphics[scale=0.3]{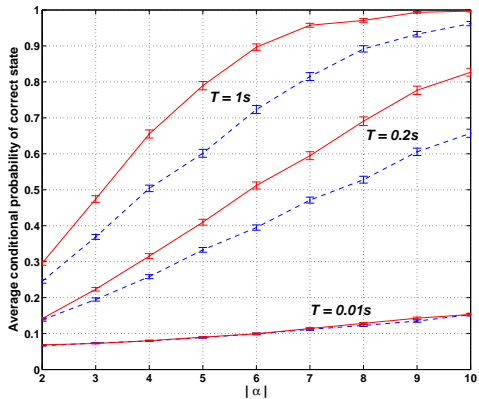}
\caption{(Color online) Contrasting the performance of the adaptive and static schemes for discriminating the $16$-element constellation of section \ref{sec:largen}. This figure shows the average \textit{a posteriori} probability of the correct hypothesis as a function of probe signal magnitude (SNR) $\alpha$ at three fixed times. The dotted (blue) lines plot results for the static scheme and the solid (red) lines plot results for the adaptive scheme. As in Fig. \ref{fig:cqedres}(b), the error bars at each data point show the standard error of the mean values plotted. \label{fig:largen_res}}
\end{figure}

Again, we compare the adaptive scheme introduced above to static heterodyne detection (here the phase was cycled at $300\pi$ rads/sec in order to be above the threshold mentioned above). This comparison is summarized in Fig. \ref{fig:largen_res}. As in Fig. \ref{fig:cqedres}.c we show the average \textit{a posteriori} probability (averaged over 500 runs, and for a particular choice of correct phase) as a function of the probe signal magnitude for three fixed times. As in the small $N$ case, the two schemes are almost indistinguishable at early times. However, for larger times the performance of the adaptive scheme becomes much better than that of the static heterodyne scheme. This example suggests that the difference in performance between the static and adaptive strategies will become more pronounced as the number of phases to distinguish becomes larger. The figure also shows that the discrimination takes longer as $N$ increases (for both the static and adaptive strategies).

Like Figure \ref{fig:cqedres}, Figure \ref{fig:largen_res} shows the results for a particular correct phase.  In Table \ref{tab:av_largen} we show the average performance of $\overline{\mathcal{P}_{corr}}$ for $|\alpha| = 5$ and two time values, averaged over all possible correct values of the phase.  Note that, as in Table \ref{tab:av_cqed}, there are two levels of averaging here, once over many runs of the discrimination task and once over the 16 possible correct states. As in the illustrative case described by Fig. \ref{fig:largen_res}, we see that the average performance of the adaptive scheme is consistently better than that of the static scheme. 

\begin{table}[htdp]
\begin{center}
\begin{tabular}{|c|c|c|c|}
\hline
Time (secs) & Detection scheme & Average $\overline{\mathcal{P}_{corr}}$ & $\sigma$ \\
\hline \hline
0.2 & Static & 0.3304 & 0.0259  \\
0.2 & Adaptive & 0.4109 & 0.0334 \\
1 & Static & 0.6195 & 0.0430 \\
1 & Adaptive & 0.7994 & 0.0442 \\
\hline
\end{tabular} 
\caption{$N=16$ detection: values of the average \textit{a posteriori} probability of the correct phase, and standard deviation of this quantity, at a short and long time instant at fixed SNR ($|\alpha | = 5$).  \label{tab:av_largen}}
\end{center}
\end{table}

\section{Conclusion}
\label{sec:conc}
We have studied the use of adaptive homodyne measurements for the task of phase discrimination. This task is especially relevant to quantum computing architectures such as circuit-QED that use harmonic oscillator modes to mediate the measurement of qubits. We have shown that adaptive measurements provide no performance advantage in the task of measuring one qubit (distinguishing between two phases). Following this, we presented numerical evidence from comparison of an adaptive homodyne strategy with a static heterodyne detection scheme that indicates an advantage in using adaptive homodyne measurements when the task is to measure more than one qubit.  This advantage is two fold: a decrease in the signal-to-noise ratio (which is essentially the magnitude of the measurement probe signal) required to perform the discrimination task to a given accuracy, and a decrease in the time required to perform the discrimination task to a given accuracy.

Although we demonstrated the utility of our scheme using the circuit-QED setup, it applies equally well to other qubit-oscillator systems. In particular, such an adaptive scheme will likely be advantageous in improving the distinguishability of the states of a single spin when measured by coupling the spin to microcavities \cite{Sug.Mac.etal-2003, Ata.Dre.etal-2007}.

One could imagine extending this study in a number of ways. The first would be to consider optimality in the $N>2$ discrimination task. This could be done by formulating a suitable differentiable performance function amenable to standard techniques of optimization, or by applying more exotic optimization techniques to a stochastic performance function. Second, the two examples studied in section \ref{sec:ng2} suggest that the advantage offered by an adaptive scheme increases as the number of phases to distinguish is increased. It would be interesting to determine exactly how the advantage (of this adaptive scheme over the best static scheme for the constellation) scales with the constellation size, $N$. Finally, we have not addressed issues of inefficient measurement here; all the above calculations have assumed unit efficiency measurement. Inefficiencies in the measurement would modify the POVM of \erf{eq:povm}, and therefore the expression for the likelihood functions and \textit{a posteriori} probabilities. We expect that measurement inefficiency will have the same effect as reducing the SNR (probe signal size, $\alpha$), however, this remains to be confirmed.

Another interesting avenue for further study is the utility of adaptive schemes for qubit state discrimination when the cavity dynamics is non-linear. Such non-linear probe devices have recently been developed \cite{Sid.Vij.etal-2006, Boa.Man.etal-2007}, and the non-linearity of the cavity results in a more responsive measurement of the qubit. An investigation of the applicability of adaptive discrimination techniques, such as the ones described in Refs. \cite{Dol-1973} and \cite{Tak.Sas.etal-2006}, to such devices would be interesting.

\section{Acknowledgements}
\label{sec:acks}
We thank the NSF for financial support under ITR Grant No. EIA-0205641, and the NSA for financial support under MOD713106A. MS would like to acknowledge useful discussions with Jay Gambetta, Alexandre Blais, and Ramon van Handel.

\bibliography{/Users/mohan/Documents/bibdesk/mybib}

\end{document}